\documentclass[pre,aps,onecolumn,superscriptaddress,showpacs]{revtex4}
\usepackage{amsmath}
\usepackage{graphicx}

\begin{document}

\title{Laser beam coupling with capillary discharge plasma\\for laser wakefield acceleration applications}

\author{G.\,A.\,Bagdasarov}
\affiliation{Keldysh Institute of Applied Mathematics, Moscow, 125047, Russia}
\affiliation{National Research Nuclear University MEPhI (Moscow Engineering Physics Institute), Moscow, 115409, Russia}
\author{P.\,V.\,Sasorov}
\affiliation{Keldysh Institute of Applied Mathematics, Moscow, 125047, Russia}
\author{V.\,A.\,Gasilov}
\author{A.\,S.\,Boldarev}
\affiliation{Keldysh Institute of Applied Mathematics, Moscow, 125047, Russia}
\affiliation{National Research Nuclear University MEPhI (Moscow Engineering Physics Institute), Moscow, 115409, Russia}
\author{O.\,G.\,Olkhovskaya}
\affiliation{Keldysh Institute of Applied Mathematics, Moscow, 125047, Russia}
\author{C.\,Benedetti}
\author{S.\,S.\,Bulanov}
\author{A.\,Gonsalves}
\author{H.-S.\,Mao}
\author{C.\,B.\,Schroeder}
\author{J.\,van\,Tilborg}
\author{E.\,Esarey}
\author{W.\,P.\,Leemans}
\affiliation{Lawrence Berkeley National Laboratory, Berkeley, California 94720, USA}
\author{T.\,Levato}
\author{D.\,Margarone}
\author{G.\,Korn}
\affiliation{Institute of Physics ASCR, v.v.i. (FZU), ELI-Beamlines Project, 182 21 Prague, Czech Republic}

\begin{abstract}
One of the most robust methods, demonstrated up to date, of accelerating electron beams by laser-plasma sources is the utilization of plasma channels generated by the capillary discharges. These channels, i.e., plasma columns with a minimum density along the laser pulse propagation axis, may optically guide short laser pulses, thereby increasing the acceleration length, leading to a more efficient electron acceleration. Although the spatial structure of the installation is simple in principle, there may be some important effects caused by the open ends of the capillary, by the supplying channels etc., which require a detailed 3D modeling of the processes taking place in order to get a detailed understanding and improve the operation. However, the discharge plasma, being one of the most crucial components of the laser-plasma accelerator, is not simulated with the accuracy and resolution required to advance this promising technology. In order to achieve good coupling of a laser pulse to the plasma channel it is necessary to have a tool for the simulation of a 3D electron distribution inside the capillary, near the open ends of the capillary, as well as near the channels supplying neutral hydrogen into the capillary. In the present work, such simulations are performed using the code MARPLE. First, the process of the capillary filling with a cold hydrogen before the discharge is fired, through the side supply channels is simulated. The main goal of this simulation is to get a spatial distribution of the filling gas in the region near the open ends of the capillary. A realistic geometry is used for this and the next stage simulations, including the insulators, the supplying channels as well as the electrodes. Second, the simulation of the capillary discharge is performed with the goal to obtain a time-dependent spatial distribution of the electron density near the open ends of the capillary as well as inside the capillary. Finally, to evaluate effectiveness of the beam coupling with the channeling plasma wave guide and electron acceleration, modeling of laser-plasma interaction was performed with the code INF\&RNO.
\end{abstract}

\pacs{
52.65.-y, 
52.65.Kj, 
52.58.Lq, 
52.38.Kd} 

\maketitle

\section{Introduction}\label{sec:intro}
Laser plasma accelerators (LPA) are leading candidates for the next generation lepton colliders and are also considered as compact electron beam sources for free-electron lasers (FEL) due to their ability to produce acceleration gradients on the order of tens to hundreds of GV/m. This ability leads to compact acceleration structures, significantly reducing the size of the corresponding facilities (see~\cite{Physics Today, LWFA_review} and references therein). Over the last 10-15 years rapid progress has been achieved in generating quasimonoenergetic electron beams by LPAs~\cite{e-beams}. In particular, the demonstration of GeV electron beams from cm-scale capillary discharge waveguides~\cite{LG}, as well as the proof-of-principle coupling of two accelerating structures, powered by different laser pulses~\cite{staging}, has increased interest in LPAs as a viable technology to be considered for compact colliders and FELs~\cite{Physics Today}.

A low current capillary discharge in resistive regime may form a radial distribution of electron density that is necessary for formation of plasma wave guide capable of transporting a channeled laser beam on a distance that is much longer than its diffraction Rayleigh range~\cite{EC,KS,KT,HS,MM,SH,BE}. This property of the capillary discharges is used in the laser plasma accelerator project BELLA~\cite{LD,LDD} based on the ideas of laser wakefield acceleration~\cite{TD,SE,GK,E3,ES2}.

The BELLA team has demonstrated recently~\cite{LG} acceleration of 4.2\,GeV electron beams containing $\sim 4\cdot10^7$ electrons per laser shot. These electron beams have 6\% energy spread and $\sim 0.3$\,mrad divergence. The electron beams were produced with the BELLA femtosecond Ti:sapphire 815\,nm laser, which gives 16\,J laser energy per 40\,fs shot focused at 52\,$\mu$m spot size with 1\,Hz repetition rate. This laser beam passed through the capillary discharge plasma that is excited preliminary with 250\,A electric current with $\sim$90\,ns duration. The current passed through the sapphire capillary of 500\,$\mu$m diameter and 9\,cm length. This capillary is prefilled with hydrogen so, after full ionization with the electric current, the axial electron density reaches $\sim 7\cdot10^{17}/\mathrm{cm}^3$. Discharge capillaries have demonstrated recently the ability to work with high repetition rate (1\,kHz)~\cite{GP}, as well as to transport and focus high energy electron beams~\cite{staging,plasma lens}. 

The theory of these capillary discharges as well as the modeling of the plasma dynamics inside them assumed usually a 1D approach (see for example~\cite{BE}). All parameters of the plasma are assumed as dependent only on time~$t$ and radius~$r$. Such an approach is relevant because the aspect ratio of capillary length $L$ and its radius $R_0$ is usually very large. For capillary discharges of this type, the skin-effect plays only a minor role so that electric field exiting the electric current is almost homogeneous across the discharge as well as the electric current density. Ampere's force ($\mathbf{j}\times\mathbf{B}$) also plays only a minor role in comparison with the plasma pressure gradient force. In this case the plasma is confined by the capillary walls. After ionization, taking of the order of 100\,ns, and plasma redistribution across the discharge, a quasi equilibrium state is established~\cite{BE}. The equilibrium implies both mechanical and thermal equilibrium. The latter one is established by the balance between Joule heating and thermal flux to the walls due to thermal conduction. The equilibrium changes adiabatically because of changing of the electric current during its pulse. As a result very simple analytic theory~\cite{BE} was developed that gives scaling laws for temperature, electron density, and their radial profiles.

However such 1D theoretical approach is insufficient to resolve several questions that are important when capillary discharge plasma is used for laser acceleration. One of them is matching and coupling of the focused laser beam with the capillary plasma wave guide. Such coupling is determined largely by the electron density distribution at the vicinity of the open end of the capillary. The distribution is determined in turn by 2D and 3D dynamics of capillary discharge plasma in this region. The plasma dynamics may be affected by the design of electrodes and other details at the open end of the capillary. Hence 3D magnetohydrodynamic (MHD) simulations of this effect should take into account the specific design. Such simulations are also important for the determination of the role of gas filling supplies on the transport of both the electron and the laser beam.

The main goal of this work is 3D MHD simulations of the capillary discharges accompanied by simulation of laser beam coupling with the capillary plasma as well as of the laser beam propagation inside the wave guide and of the process of injected electron bunch acceleration. Parameters of the discharge will be similar to that used in the BELLA experiment~\cite{LG,staging}. We are interested mainly in the electron density distribution and its dynamics in the vicinity of the capillary open ends. Design of the capillary in this region is formed by both metal electrodes and dielectrics. This complex structure should be taken into account to determine the time-dependent electric current density distribution, which in turn determines ohmic heating of the discharge plasma and, as a result, its dynamics.

Our preliminary simulations showed that the initial distribution of neutral hydrogen near the capillary orifice before the electric current pulse influences significantly the resulting plasma flowing out of the capillary orifice. For this reason we performed simulations on initial filling of the capillary with neutral hydrogen, using 3D gas-dynamic simulations that take into account the capillary design including filling supplies. Results of the gas simulations were used as an initial condition for the capillary discharge MHD simulations. The code MARPLE~\cite{GB} was used for both types of simulations.

Modeling of the laser-plasma interaction was done with the code INF\&RNO (INtegrated Fluid \& paRticle simulatioN cOde)~\cite{INFRNO}. INF\&RNO allows simulation of laser evolution and electron bunch dynamics in the vicinity of the open end of the capillary as well as inside it. As a result we are able to evaluate the effectiveness of the whole process of laser beam coupling in the plasma channel and the acceleration of electrons.

The paper is organized as follows. In section~\ref{sec:code} we describe the architecture of the MHD code MARPLE. In section~\ref{sec:filling} we show the results of the capillary filling with a neutral gas and the resulting density profile before firing the discharge. Results of the capillary discharge modeling and the consequent plasma density profiles are discussed in section~\ref{sec:discharge}. In section~\ref{sec:comparison} we compare the results of the 3D MHD simulations of the capillary discharge with the results obtained by 1D MHD code NPINCH. We use the simulated 3D plasma profile as initial condition for PIC simulations and report the results in section~\ref{sec:lwfa}. Conclusions are drawn in section~\ref{sec:outro}. 

\section{3D MHD code MARPLE}\label{sec:code}

MARPLE (Magnetically Accelerated Radiative PLasma Explorer)~\cite{GB} is an Eulerian numerical tool designed for simulations of radiative MHD problems related to experiments with magnetically driven high energy density plasmas (HEDP) and pulsed power energetic (PPE). This is an object-oriented parallel code (using MPI and CUDA techniques) designed for scientific simulations on systems performing distributed computations.

The code works with irregular unstructured meshes whose cells may be tetrahedrons, triangular or quadrilateral prisms, as well as four-sided pyramids. Such type of meshing provides higher flexibility and a possibility of mesh adaptation to the problem peculiarities. The computer model is based on additive accounting of physical processes. The resulting approximation scheme is homogeneous (as to the treatment of shocks and other types of discontinuities) and conservative. The overall time-advance procedure is a second-order predictor-corrector. The parallel computations are implemented via domain decomposition. A new feature implemented in the code is that the ghost (often referred to as "halo") elements are also used for setting various symmetry boundary conditions associated with the problem under study. Translational, rotational, mirror symmetries and their combinations are extensively used.

Agile architecture of the code MARPLE allows easily adapt it for the simulation of capillary discharges that are characterized by significantly lower densities and temperatures compared to HEDP and PPE problems, as well as specific matter properties.

Physical model implemented in MARPLE is based on two-temperatures (ion and electron), one-fluid MHD model including various essential elements such as electron-ion energy relaxation, energy dissipation and radiative transfer:
\begin{align}
  &\partial_t\rho + \nabla\left(\rho \mathbf{v}\right) = 0,\label{eq:1}\\
  &\rho\left(\partial_t + \mathbf{v}\cdot\nabla\right)\mathbf{V} =
    -\nabla p + \mathbf{j}\times\mathbf{B}/c,\label{eq:2}\\
  &\rho\left(\partial_t + \mathbf{v}\cdot\nabla\right)\varepsilon_e + p_e\nabla\mathbf{v}
    = \nabla\left(\kappa_e \cdot \nabla T_e\right) - Q_{ei} + \mathbf{j}^2/\sigma - Q_r,\label{eq:3}\\
  &\rho\left(\partial_t + \mathbf{v}\cdot\nabla\right)\varepsilon_i + p_i\nabla\mathbf{v}
    = \nabla\left(\kappa_i \cdot \nabla T_i\right) + Q_{ei},\label{eq:4}\\
  &\partial_t\mathbf{B} + \nabla\times\left(\nu_m \cdot \nabla\times\mathbf{B}\right) =
    \nabla\times\left(\mathbf{v}\times\mathbf{B}\right),\label{eq:5}\\
  &\mathbf{j} = c/4\pi \cdot \nabla\times\mathbf{B},\quad\nu_m =
    c^2 / 4\pi\sigma.\label{eq:6}
\end{align}
Here we use standard designations for magnetic field strength ($\mathbf{B}$), plasma density ($\rho$), its velocity ($\mathbf{v}$), electron and ion temperatures ($T_{e,\,i}$), electron and ion specific energies ($\varepsilon_{e,\,i}$) and pressures ($p_{e,\,i}$). These equations contain also dissipative coefficients: $\kappa_{e,\,i}$ and $\sigma$ which describe electron and ion thermal conductivities, and electric conductivity, respectively. Another important element of the model is matter properties. They are determined in the frame of this model by density and temperature dependencies of the values mentioned above: $\varepsilon_{e,\,i}$, $p_{e,\,i}$,  $\kappa_{e,\,i}$ and $\sigma$. In the present work we investigate the discharge dynamics in a capillary prefilled by hydrogen. We adopt here properties of the hydrogen plasma described in~\cite{BE}. The model assumes in particular local thermodynamic equilibrium (LTE) for ionization degree of moderate and fully ionized hydrogen plasma. The energy cost of ionization is included  in the specific energy of electrons, $\varepsilon_e$. The degree of ionization determines $p_e$ and $\varepsilon_e$. Electric conductivity, $\sigma$, as well as electron thermal conductivity $\kappa_e$ are determined by collisions of electrons with both hydrogen ions and neutrals. The model~\cite{BE} includes also energy exchange rate $Q_{ei}$ between electrons and ions (neutrals). We may neglect radiation losses $Q_r$ in \eqref{eq:3} for the nearly-fully ionized hydrogen plasma. The model adopted for the present simulations is capable to describe relatively long process of plasma ionization, when we start from weakly ionized hydrogen.

The model cannot describe the initial electric breakdown of the cold hydrogen. Simulation of this process requires to take into account non-stationary kinetics of ionization and excited atomic and molecular states populations, and violation of quasi-neutrality. These processes, especially the latter one, cannot be included in the model~\eqref{eq:1}-\eqref{eq:6}. However, the breakdown lasts a  short time, only about several nanoseconds, and presumably does not affect the well developed plasma discharge considered in this paper.

\section{Simulation of capillary filling}\label{sec:filling}

Simulation of a capillary discharge was split into two stages. First, we simulate the process of filling of the capillary with cold hydrogen before the discharge is initiated. The main goal of this stage is to get a spatial distribution of the gas in the region near the open ends of the capillary needed for the second stage~-- capillary discharge simulation.

We have used the same physical model~\eqref{eq:1}-\eqref{eq:6} for both stages, but for the first one we exclude the magnetic field and ionization, as well as the electron's temperature from the equations (hence it is just one-temperature ideal hydrodynamic system with thermal conductivity taken into account). The full system of equations~\eqref{eq:1}-\eqref{eq:6} is solved for the next stage.

\begin{figure}[h!t]
  \includegraphics[width=0.75\textwidth]{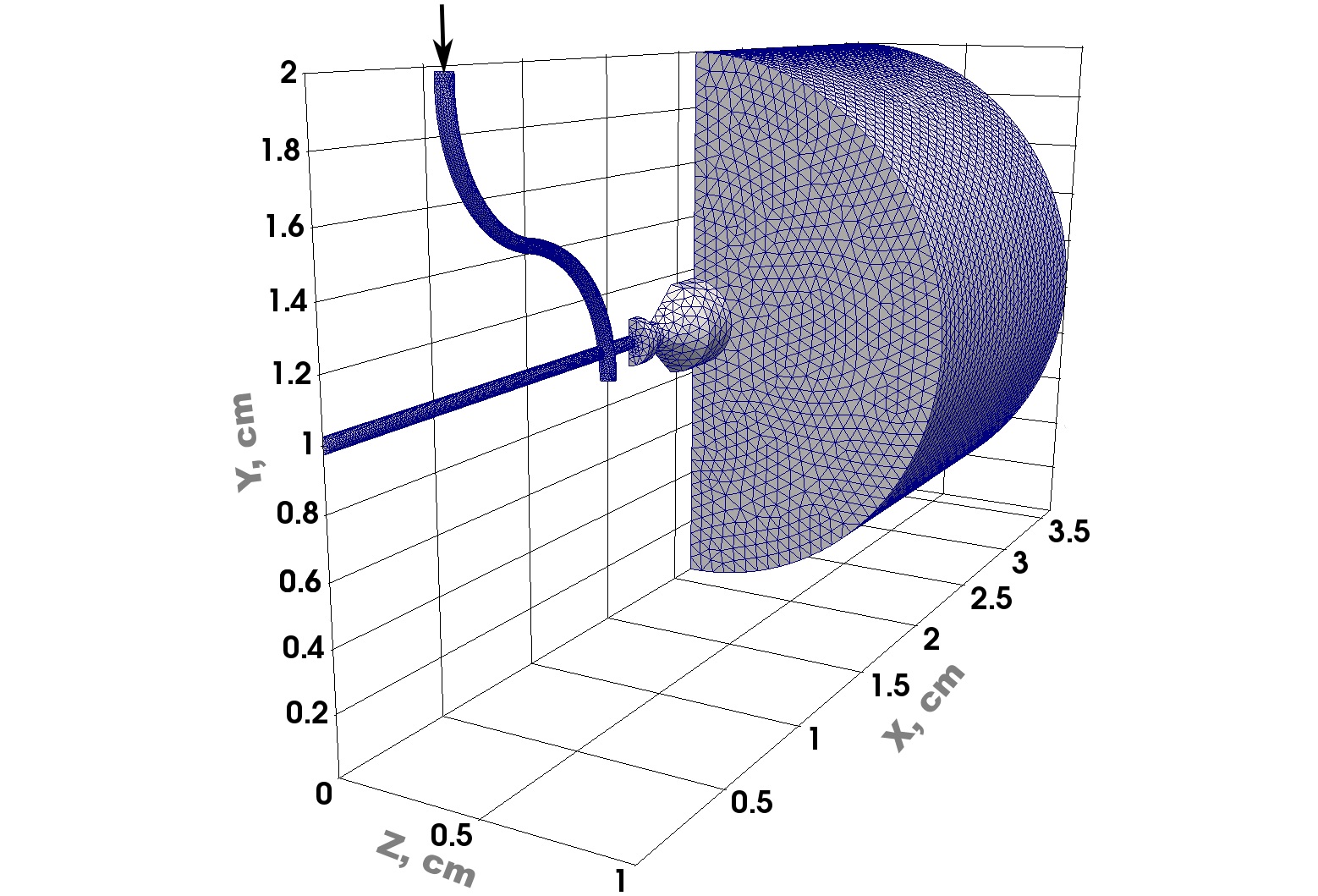}
  \caption{Computational domain with its discretization for the problem of filling. Filling supply is shown with the arrow.\label{fig:cap0init}}
\end{figure}

The geometrical parameters of the capillary as well as the initial and boundary conditions correspond to some experiments from~\cite{LG,staging}: sapphire capillary with length of 3.3\,cm and inner diameter of 0.05\,cm was used for the simulation. The real geometry was taken into account for this stage, including the insulators, the supplying channels as well as the electrodes (see Fig.\,\ref{fig:cap0init}).

Supplying channels are joined to the reservoir filled with hydrogen at pressure $p$~= 40\,Torr and at room temperature. The initial state inside the capillary is vacuum and at the moment $t$~= 0 the gas from reservoir begins to flow into the capillary through supplying channels (arrow in Fig.\,\ref{fig:cap0init}). In order to avoid the necessity of modeling the gas flow in the whole reservoir and supplying system, special boundary conditions were set at the inlet boundary (arrow in Fig.\,\ref{fig:cap0init}). These conditions can be expressed as the equality of the entropy $S$ and full enthalpy $h=\varepsilon + p/\rho +\mathbf{v}^2/2$ to the values corresponding to the stagnant gas at the given state (i.\,e. $p$ = 40\,Torr and $T$ = 293\,K). Using symmetric boundary conditions, we simulate the first stage only in a quarter of the full geometry. Final discretization with 120 thousand of tetrahedral cells is shown at Fig.\,\ref{fig:cap0init}.

\begin{figure}[h!t]
  \includegraphics[width=0.95\textwidth]{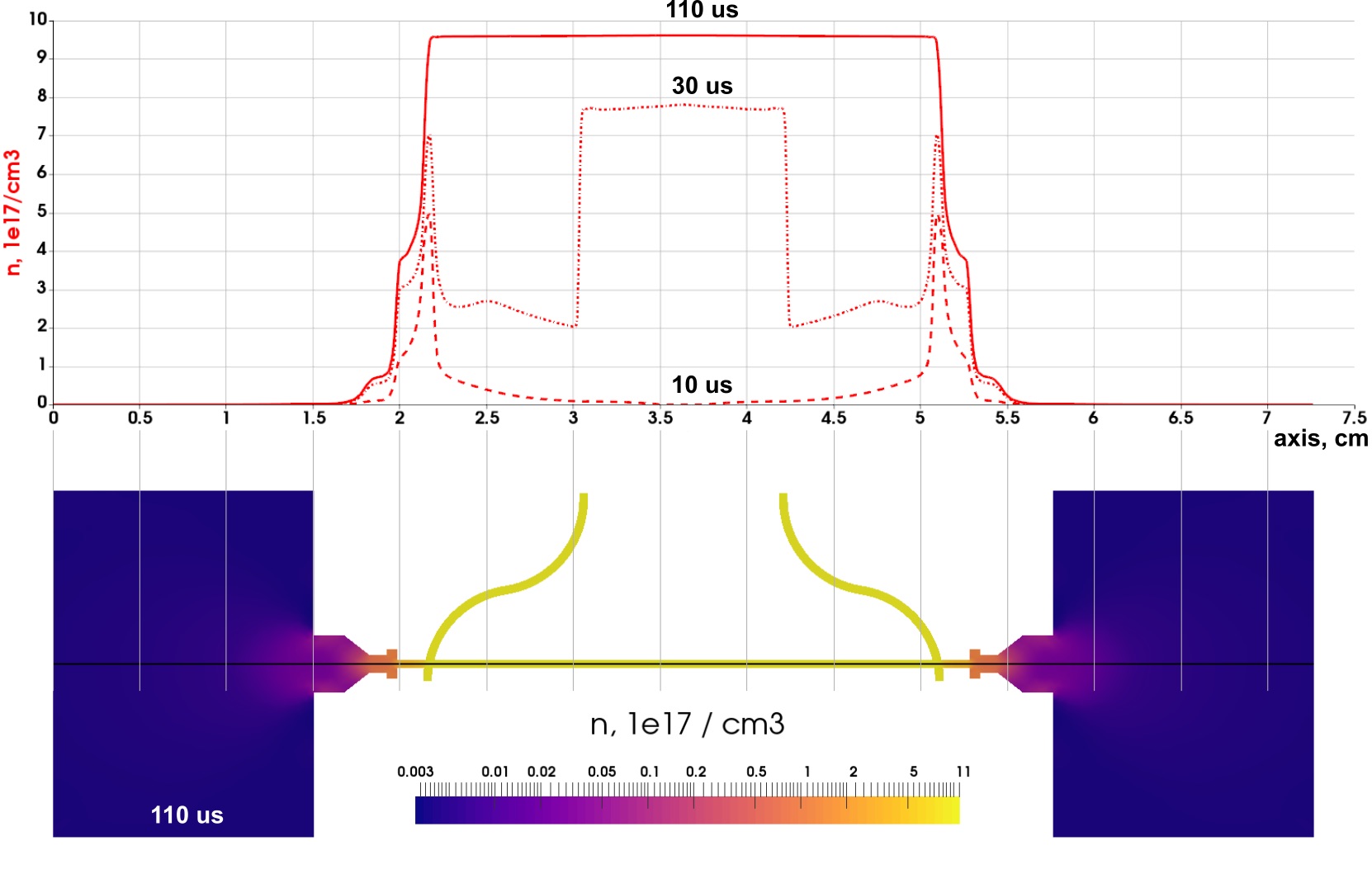}
  \caption{On the top frame density distribution along the capillary axis (black line on the bottom frame) at the moments $t$~=~10, 30 and 110\,$\mu$s are shown. Density distribution inside the capillary and its open ends as well as in supplying channels at the moment $t$~= 110\,$\mu$s (steady state stage) is presented on the bottom frame.\label{fig:cap0n}}
\end{figure}

Fig.\,\ref{fig:cap0n} shows the hydrogen distribution in the computational domain. The top frame of the figure shows the hydrogen distributions along capillary axis at the moments $t$~= 10, 30 and 110\,$\mu$s. The bottom frame shows the spatial distribution of hydrogen inside the supplying channels, capillary and near its open ends when the solution becomes steady (at 110\,$\mu$s). It is seen from Fig.\,\ref{fig:cap0n} that the gas is distributed uniformly deeply inside the capillary while near its open ends the distribution becomes non-uniform. The position of the start of this non-uniformity coincides with the position of the ends of the supplying channels inside the capillary. We will see that the spatial scale of the non-uniform distribution of electron density along the capillary axis in the vicinity of the open ends of the capillaries at the quasi-stationary stage of the discharge is significantly larger than the analogous spatial scale in the problem of filling, at least for the present geometry. Fig.\,\ref{fig:cap0n} shows also that the gas jet extends a distance of about 7\,mm from the open ends of the capillary. We will see in the next section that this distance is larger than the length of a portion of the gas jet, which is ionized during the discharge. For this reason, simulations of filling process are necessary to get valid electron density distribution outside the capillary.

\section{Simulation of capillary discharge}\label{sec:discharge}

The main goal of this stage is to obtain a time-dependent spatial distribution of the electron density near the open ends of the capillary as well as inside the capillary. The gas-dynamics simulations determined the initial hydrogen density distribution for the discharge simulation. We used the complete physical model~\eqref{eq:1}-\eqref{eq:6} without any exclusions to simulate this stage.

\begin{figure}[h!t]
  \includegraphics[width=0.9\textwidth]{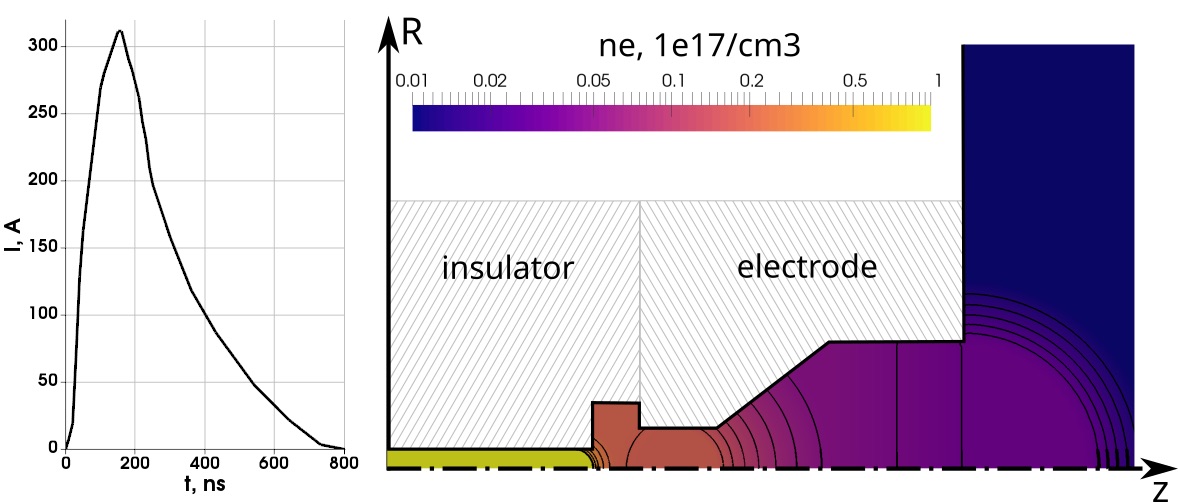}
  \caption{Computational domain for the second stage simulation on the right: simplified geometry and initial hydrogen distribution are shown. Shaded regions show the corresponding boundaries. The experimental electric current profile is presented on the left.\label{fig:cap1init}}
\end{figure}

A slightly simplified geometry of the capillary was used for this stage~-- notably the supplying channels were removed from the computational domain (see Fig.\,\ref{fig:cap1init}). The main reason is that taking into account the supplying channels at this stage requires much more detailed discretization. Indeed the nontrivial geometry of the capillary leads to the fact that lines of magnetic field cross the interface between discharge plasma and insulator, and thus does not allow us to setup a proper boundary condition on that interface. Therefore we were forced to simulate the magnetic field distribution in the insulator together with the MHD simulation of the discharge plasma, which leads to a significant expansion of the computational domain. Despite this simplification we also need much more detailed discretization to obtain reliable numerical results for a discharge simulation (50 cells along capillary radius are required for the second stage instead of just 5 for the first one). The estimated size of the computational mesh for the second stage,  taking into account the supplying channels, is about $10^8$ cells. However if we ignore the supplying channels, then the problem gains rotational symmetry about the axis of the capillary, allowing significant reduction in the computational domain (1/12 of the full domain instead of 1/4 for the first stage). In this paper we consider the simpler case without taking into account the supplying channels on the discharge stage.

The hydrogen density distribution obtained from the first stage of the discharge simulation was adapted to the simplified geometry and used as initial conditions for the second stage. The current pulse parameters are taken from the experiment~\cite{LG}, and its profile is shown at Fig.\,\ref{fig:cap1init}. Initially there is no current inside the channel and hence, in order to initiate the discharge, the hydrogen is assumed to be slightly ionized ($T_e$ = $T_i$ = 0.5\,eV everywhere) at the beginning of the second stage. The simulation time is within the range, $t\in\left[0, 250\right]$\,ns.

Appropriate boundary conditions were set on the insulator and electrode boundaries (see Fig.\,\ref{fig:cap1init}). The normal component of plasma velocity is zero in both cases. We set constant temperatures of electrons and ions ($T_e$ = $T_i$ = 0.5\,eV) at the insulator boundary that are much less than typical plasma temperature in the discharge. The azimuthal component of magnetic field at the insulator boundary is set to $B_\phi (R(z),t) = 2I/(c\cdot R(z))$, where $I(t)$ is experimental total electric current through the discharge, and $R(z)$ is radius of simulation domain. We set to zero the tangent component of electric field at the electrode boundary, such that $(\nabla\times\mathbf{B})_{||} = 0$ in equations~\eqref{eq:1}-\eqref{eq:6}.

\begin{figure}[h!t]
  \includegraphics[width=0.95\textwidth]{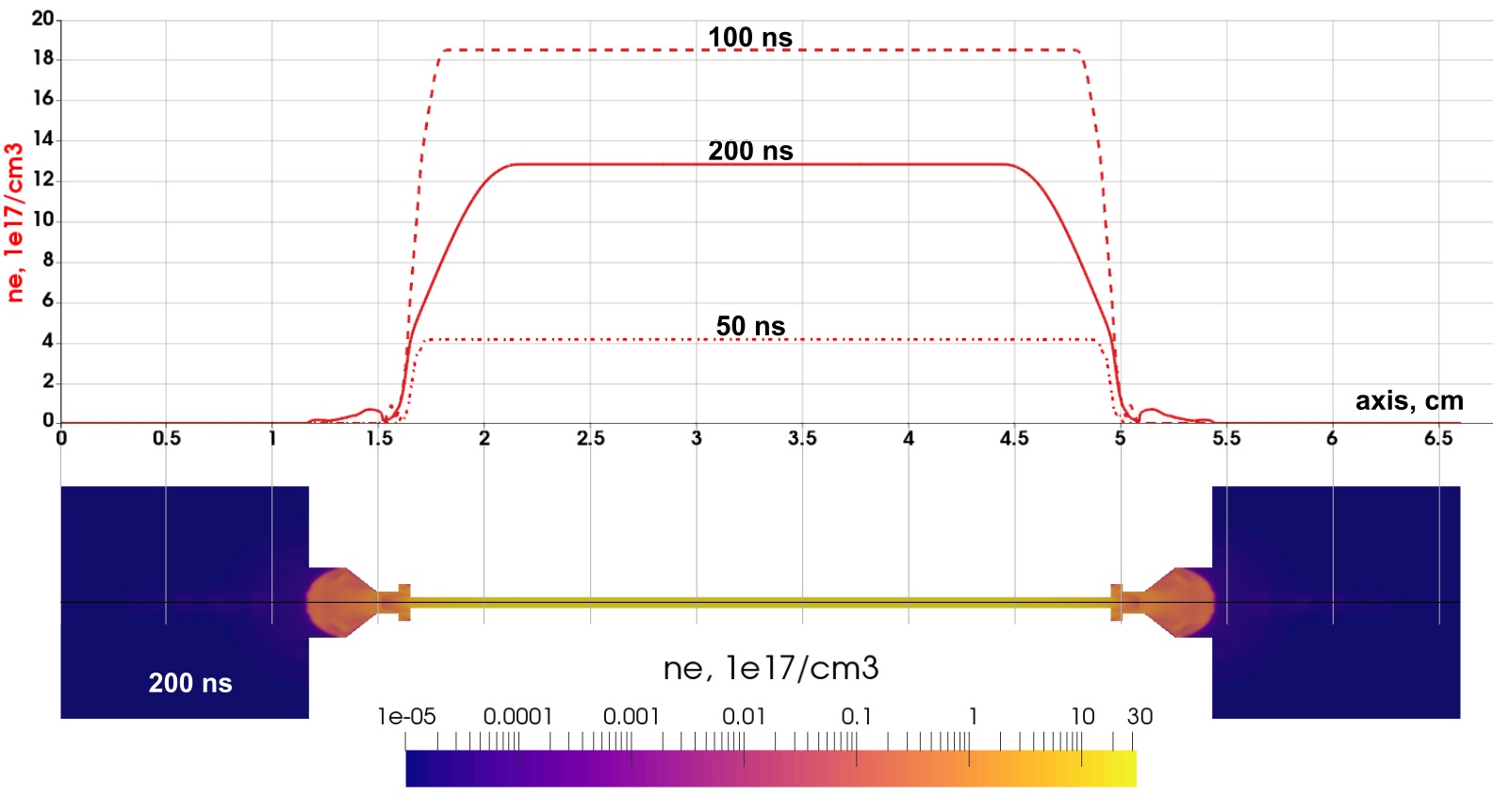}
  \caption{Electron density distributions along capillary axis at several moments (top). Spatial distribution of electron density in the capillary and near its open ends at $t$ = 200\,ns after discharge is started (bottom).\label{fig:cap1ne}}
\end{figure}

\begin{figure}[h!t]
  \includegraphics[width=0.95\textwidth]{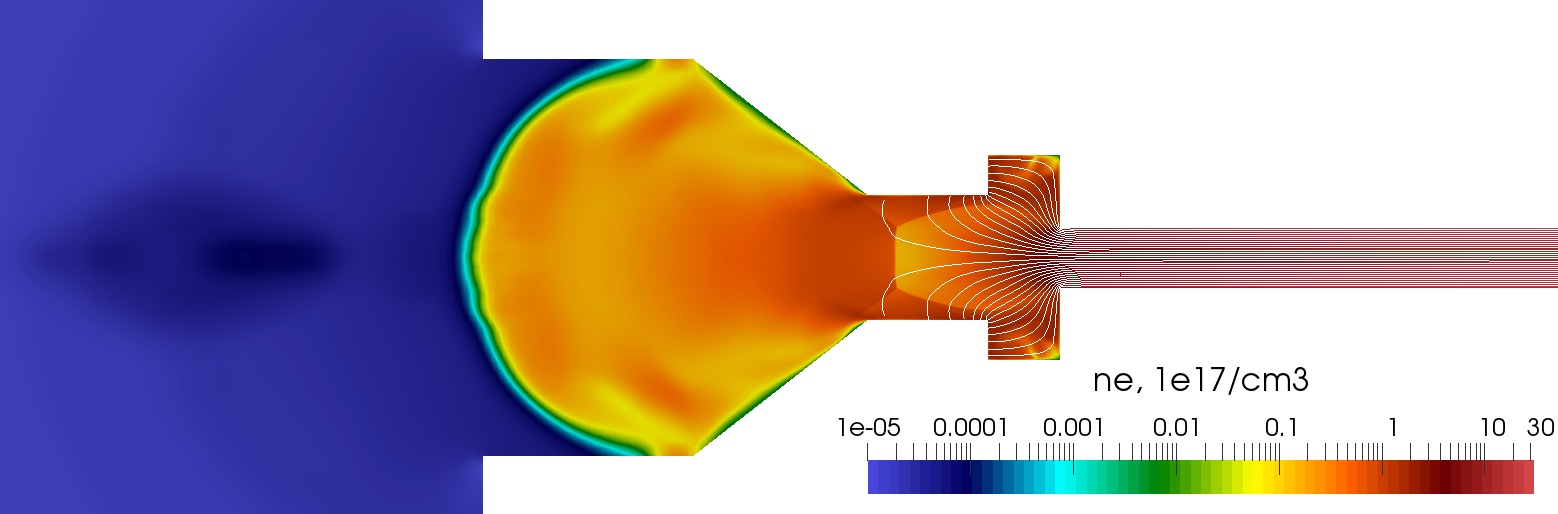}
  \caption{Electron density distribution near the open end (color) together with the streamlines of current density (white lines) at $t$ = 200\,ns after discharge is started.
\label{fig:cap1nej}}
\end{figure}

\begin{figure}[h!t]
  \includegraphics[width=0.95\textwidth]{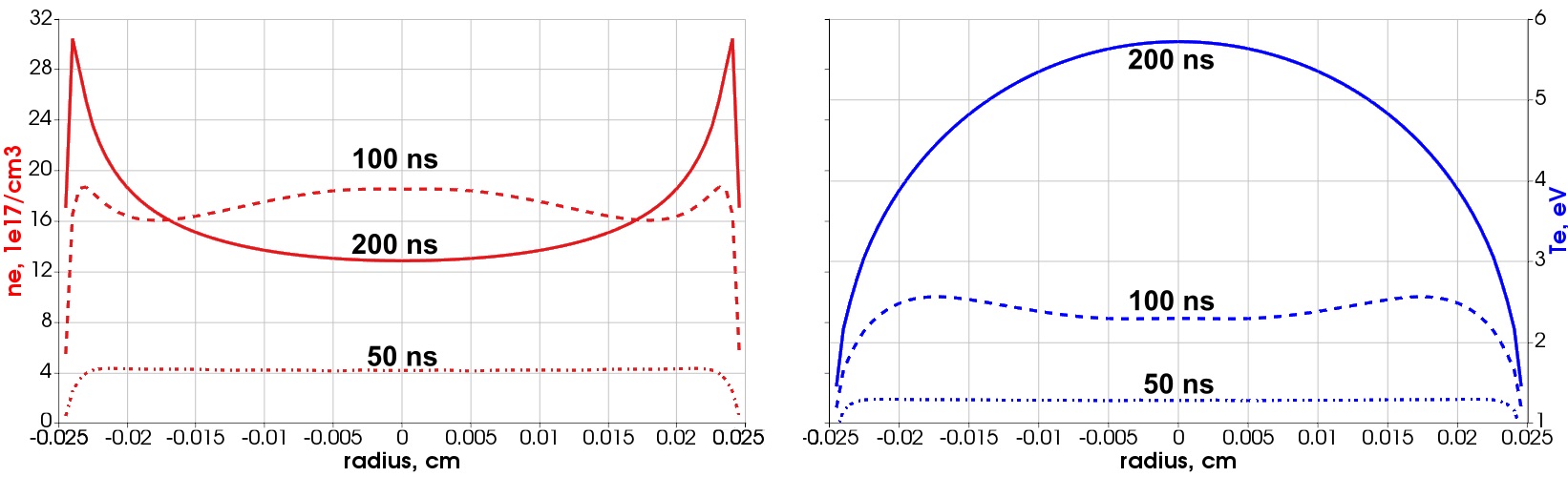}
  \caption{Radial electron density (left) and temperature (right) distributions inside the capillary at several moments of the discharge simulation.\label{fig:cap1neTe}}
\end{figure}

Simulation results of the second stage are presented in Figs.\,\ref{fig:cap1ne}-\ref{fig:cap1neTe}. The electron density spatial distribution at 200\,ns after discharge (slightly after current reached its maximum) as well as its axial distribution at several moments are shown at Fig.\,\ref{fig:cap1ne}. Complex structure of the plasma outflow from the capillary at the same moment of time is shown in the Fig.\,\ref{fig:cap1nej} (here we used another color palette for better visualization of the structure). On the same figure, streamlines of the current density are presented. These streamlines show that the boundary conditions are correct. Plots of electron density and temperature distributions along capillary diameter far away from its open ends at several moments are presented at Fig.\,\ref{fig:cap1neTe}.

The simulated plasma parameters deep inside the capillary are in a good agreement with both experimental and numerical results of other simulations (see Sec.\,\ref{sec:comparison}). We may conclude that the described physical model~\eqref{eq:1}-\eqref{eq:6} and its implementation in the code MARPLE are suitable for this kind of simulations. Fig.\,\ref{fig:cap1ne} shows that the complex geometry of the electrode (even without taking into account the supplying channel) leads to a complex flow pattern with direct and reflected shock waves near the open end of the capillary, which may affect the laser beam focusing at the entrance of the capillary.

We see that the spatial scale of ionized plasma jet outside the capillary is significantly less than the axial size of the jet in the problem of filling (see Sec.\,\ref{sec:filling}). Hence results of the problem of filling affects the structure and size of the plasma jet at the stage of discharge.

We see also that axial electron density in the capillary is suppressed significantly at the distance from the capillary end of about 5\,mm. This effect may be important for coupling of laser beam with the capillary plasma wave guide.

\section{Comparison with the 1D simulations}\label{sec:comparison}

To have an additional verification for the 3D code MARPLE we performed special joint simulations with the 1D code NPINCH~\cite{lpb}. The code NPINCH is based on a physical model that differs from the physical model described in Sec.\,\ref{sec:code} by taking into account all effects of electron magnetization. These effects play a negligible role for the capillary discharges considered here. We checked preliminary that the code NPINCH reproduces exactly all results from~\cite{BE} concerning the discharge plasma. All numerical results from~\cite{BE} were obtained with the code PICA~\cite{PICA}. For the thermodynamical completion of the physical model used in the code NPINCH, the same model for hydrogen plasma properties as in Sec.~\ref{sec:code} was used. In addition to the dimensionality, the code MARPLE is three-dimensional, whereas the code NPINCH is one-dimensional, the codes differ strongly by the mesh realizations of Eqs.~\eqref{eq:1}-\eqref{eq:6}. We used the code NPINCH for simulation of the same discharge that has been described in Sec.\,\ref{sec:discharge}.

\begin{figure}[h!t]
  \includegraphics[width=0.95\textwidth]{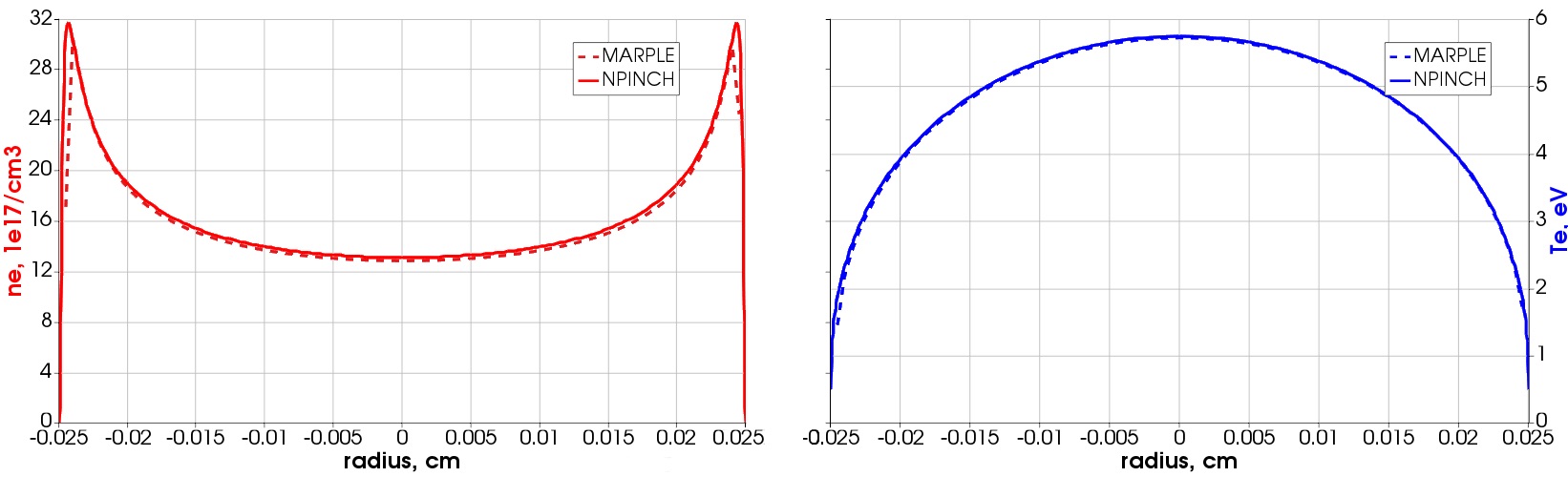}
  \caption{Comparison of radial electron density (left) and temperature (right) distributions obtained using codes NPINCH (solid) and MARPLE (dashed).\label{fig:cap1neTe_compare}}
\end{figure}

The parameters of the discharge plasma simulated deeply inside the capillary with the code MARPLE coincide with a good accuracy with the plasma parameters simulated with the code NPINCH (see Fig.\,\ref{fig:cap1neTe_compare}). The radial distributions of electron density and temperature are shown in Fig.\,\ref{fig:cap1neTe_compare}. The differences are small and caused mainly by the difference in mesh realizations and tend to 0, when all mesh steps tend to 0.

\section{Simulation of the laser wakefield acceleration}\label{sec:lwfa}

In what follows we use the simulated plasma density profile as an input for the Particle-in-Cell (PIC) simulations of the electron acceleration by high intensity laser pulse, following the setup used in recent BELLA center experiments~\cite{LG}. Modeling of the laser-plasma interaction was done with the code INF\&RNO (INtegrated Fluid \& paRticle simulatioN cOde). INF\&RNO is a code based on reduced physics models specifically designed to efficiently simulate laser-plasma accelerators.  It addresses the need for extensive numerical modeling by carefully selecting the amount and type of physics to compute. The code works in 2D cylindrical (r-z) geometry, and makes use of the averaged ponderomotive force approximation to describe the interaction of the laser pulse with the plasma. The adoption of the cylindrical geometry allows the description of key 3D physics (laser evolution, electromagnetic field structure) at 2D computational cost. The code features an improved laser solver, which retains the full wave operator, and enables an accurate description of the laser pulse evolution deep into depletion even at a reasonably low resolution. The plasma can be modeled using either a PIC or a fluid description. Both PIC and fluid modalities are integrated in the same computational framework. Besides the conventional time-explicit modality, it also features a quasi-static PIC/fluid modality. The basic numerical features of the INF\&RNO framework are discussed in detail in~\cite{INFRNO}. Compared to conventional 3D finite-difference time-domain PIC codes, INF\&RNO allows for a speedup of several orders of magnitude (between 2 and 5, depending on the particular problem and numerical settings) in the calculation time, still retaining physical fidelity. 
\begin{figure}[h!t]
  \includegraphics[width=0.45\textwidth]{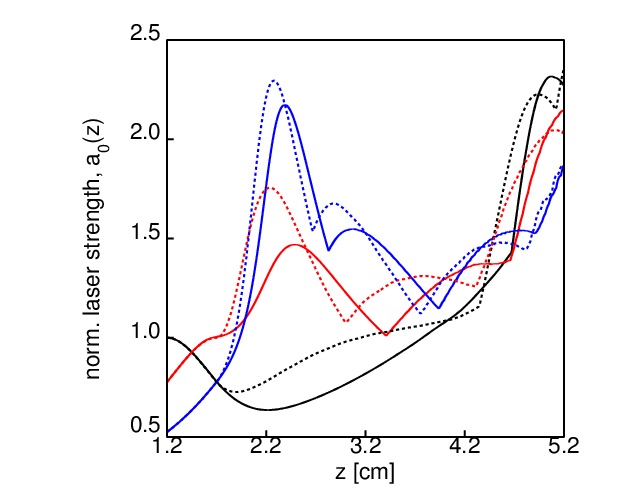}\includegraphics[width=0.45\textwidth]{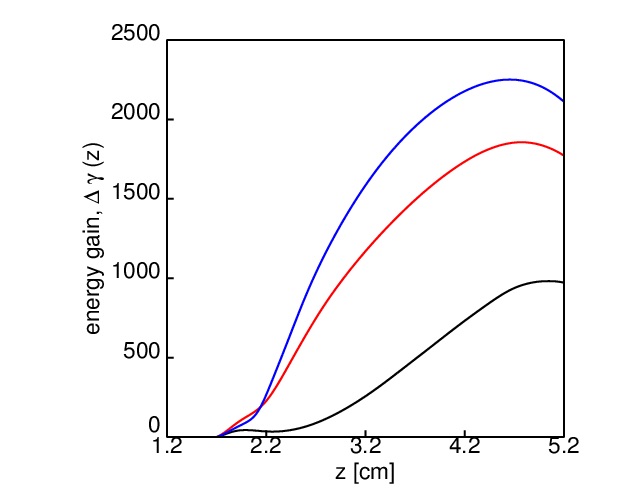}
  \caption{(a) The evolution of the normalized laser field strength as a function of the propagation distance for 3 values of the focus position $z_f=1.2$\,cm~(black), $z_f=1.7$\,cm~(red), $z_f=2.2$\,cm~(blue), the dashed lines correspond to the longitudinally rectangular density profile, $n_e(z)=n_e(z=3 ~\text{cm})$ for $1.7 ~\text{cm}<z<4.7~\text{cm}$,  $n_e(z)=0$ for  $z>4.7~\text{cm}$ and $z<1.7~\text{cm}$; (b) the evolution of the energy gain ($\Delta \gamma$) as a function of the propagation distance for 3 values of the focus position $z_f=1.2$\,cm~(black), $z_f=1.7$\,cm~(red), $z_f=2.2$\,cm~(blue).\label{fig:LWFA}}
\end{figure}

The simulations were performed for the laser pulse with $a_0=1$, where $a_0$ is the normalized amplitude of the electromagnetic field vector-potential, $a_0=eA/m_e c$. The laser pulse duration, $L$, was chosen to satisfy the relation $k_p L=2$, i.e., the resonant plasma wave excitation, $k_p=2\pi/\lambda_p=[4\pi e^2 n_e/m_e c^2]^{1/2}$, where $\lambda_p$ is the plasma wavelength, and $n_e=1.3\cdot10^{18}\,\mathrm{cm}^{-3}$ is the on-axis plasma density. The width of the laser pulse is $k_p w_0=8.6$ (corresponding to $w_0=40\,\mu$m), where $w_0$ is defined as the radius at which the intensity drops to $1/e^2$ of the peak value. The ratio $P/P_c=2.33$, where $P$ is the power of the laser pulse and $P_c=17(\omega/\omega_p)^2$\,GW is the critical power for relativistic self-focusing is low enough to avoid strong self-focusing effects. We note that the matched radius for the simulated density profile is $R_{matched}=58\,\mu$m, which means that in the limit $a_0\ll 1$ the laser pulse with $w_0=40\,\mu$m would oscillate radially while propagating in such a channel. However the self focusing compensates to some degree these oscillations, allowing for the stable propagation of the laser pulse in the plasma channel. The evolution of the laser pulse amplitude is shown in Fig.\,\ref{fig:LWFA}a for three cases of laser focus position: $z_f=1.2$\,cm (black), $z_f=1.7$\,cm (red), and $z_f=2.2$\,cm (blue). While the case with the laser focused at the entry of the capillary expands as it begins to propagate inside the plasma channel, and only towards the end of the capillary is focused, in the other two cases the laser is focused near 1\,cm inside the capillary and then oscillate as they propagate forward. In order to compare the effect of the density up-ramp and down-ramp at the entry and exit of the capillary, we show also in Fig.\,\ref{fig:LWFA}a the evolution of the laser pulse amplitude in the case of a longitudinally rectangular density profile: $n_e(z)=n_e(z=3 ~\text{cm})$ for $1.7 ~\text{cm}<z<4.7~\text{cm}$,  $n_e(z)=0$ for  $z>4.7~\text{cm}$ and $z<1.7~\text{cm}$ (dashed curves). One can see a significant difference in the maximum field amplitude, which indicates different evolution of the plasma wave, and thus potentially different result for the accelerated electron bunch energy.      
\begin{figure}[h!t]
 \includegraphics[width=0.5\textwidth]{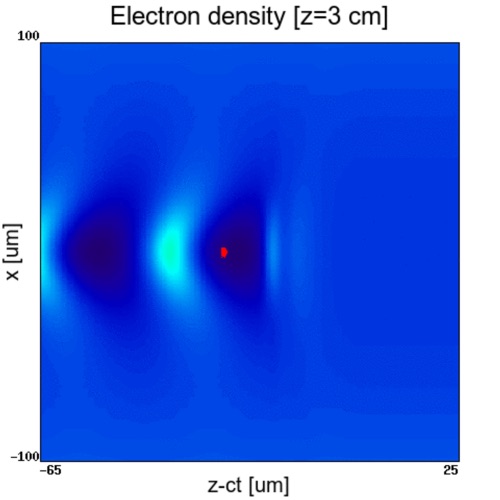}
  \caption{A snapshot of the plasma wave with the accelerating electron bunch (in red)\label{fig:LWFA1}}
\end{figure}

The electron bunch is externally injected and is initially located at the edge of the defocusing domain. Even for $z_f=1.2$ the bunch is not lost, which could have been expected due to the evolution of the plasma wave, but the energy gain is small since the laser intensity remains low. The case $z_f=2.2$\,cm shows the max energy gain of about 1.1\,GeV in 3.5\,cm, as shown in Fig.\,\ref{fig:LWFA}b. We note that for the density of the channel considered here, the dephasing length is around 3\,cm (see the electron density distribution and accelerated electron bunch in Fig. \ref{fig:LWFA1} at dephasing), which is clearly seen in the dependence of the energy gain on the distance in Fig. \ref{fig:LWFA}b.

\section{Discussion and Conclusions}\label{sec:outro}

We demonstrated that the 3D MHD code MARPLE combined with the use of modern supercomputers is able to simulate time-dependent spatial distribution of electron density in the vicinity of open ends of the capillary discharge used in the BELLA project. 

We showed that simulation of the process of initial pre-filling of the capillary with neutral hydrogen is necessary for obtaining valid results for electron density distribution in vicinities of capillary open ends during the discharge. The electron density distribution depends on extended distribution of the neutral hydrogen in this region before discharge initiation. The process of pre-filling lasts about $10^3$ times longer than the current rise time of the discharge.

We see that a complex structure of shock waves is formed in the vicinity of the open ends of the capillary during the discharge. It is important for the coupling of the laser beam and plasma wave guide in the capillary and for trapping of electrons in the process of acceleration. We plan to investigate in forthcoming papers how the shock wave structure depends on a design of junction of the capillary and electrodes.

We see also that deep inside the capillary the radial distributions of the electron density coincides well with what is obtained using the 1D model~\cite{BE}. Nevertheless a certain part of the capillary discharge at a distance of about 5\,mm from the ends of the capillary is affected by the existence of the open ends. The axial electron density of the discharge in this region is significantly lower than the axial electron density deep inside the capillary. We see that a typical spatial scale of the lower electron density region is significantly larger than the distance between the open ends of supplying channels inside the capillary and the open ends of the capillary itself.

We performed simulations of the laser wakefield acceleration of an electron bunch, using the plasma density profile from the MHD simulations. The results showed the dependence of the laser pulse evolution and electron energy gain on the distance, at which the laser pulse is focused inside the capillary. In order to be able to accurately simulate this dependence the profile of the plasma outflow from the channel ends is needed, which is supplied by the MHD simulation results. As we showed in comparing the MHD simulated density profile and a profile without the plasma outflow the laser pulse evolution for these two profiles is different, which implies the necessity for the accurate modeling of the plasma density profile variation due to shape of the capillary. The coupling of the 3D MHD code to the PIC code provides the first step to the self-consistent start-to-end simulation of the laser plasma accelerator.

\section*{ACKNOWLEDGMENTS}

The work was supported in part by the Russian Foundation for Basic Research (Grant No.\,15-01-06195), by the Competitiveness Program of National Research Nuclear University MEPhI (Moscow Engineering Physics Institute), contract with the Ministry of Education and Science of the Russian Federation No.\,02.A03.21.0005, 27.08.2013 and the basic research program of Russian Ac. Sci. Mathematical Branch, project 3-OMN RAS. The work at Lawrence Berkeley National Laboratory was supported by US DOE under contract No.\,DE-AC02-05CH11231. This research was also sponsored by the project ELI~-- Extreme Light Infrastructure–Phase\,2
(CZ.02.1.01/0.0/0.0/15\_008/0000162) through the European Regional Development Fund and by the Ministry of Education, Youth and Sports of the Czech Republic (National Program of Sustainability II Project No. LQ1606).

\end{document}